\documentclass[aps,prx,showpacs,twocolumn,superscriptaddress]{revtex4-1}

\usepackage[pdftex]{graphicx} 
\usepackage{natbib}
\usepackage{booktabs}
\usepackage{color}
\usepackage{amsfonts}

\newcommand{\one}{Fig.~\ref{f1}}
\newcommand{\two}{Fig.~\ref{f2}}
\newcommand{\three}{Fig.~\ref{f3}}
\newcommand{\four}{Fig.~\ref{f4}}
\newcommand{\five}{Fig.~\ref{f5}}
\newcommand{\ones}{Fig.~\ref{fs1}}
\newcommand{\twos}{Fig.~\ref{fs2}}
\newcommand{\threes}{Fig.~\ref{fs3}}
\newcommand{\fours}{Fig.~\ref{fs4}}
\newcommand{\fives}{Fig.~\ref{fs5}}
\newcommand{\sixs}{Fig.~\ref{fs6}}

\newcommand{\mote}{MoTe$_2$} 
\newcommand{\wte}{WTe$_2$}

\begin{document}
\title{Fermi arcs and their topological character in the candidate type-II Weyl semimetal \mote{}}

\author{A. Tamai}
\affiliation{Department of Quantum Matter Physics, University of Geneva, 24 Quai Ernest-Ansermet, 1211 Geneva 4, Switzerland}
\author{Q. S. Wu}
\affiliation{Theoretical Physics and Station Q Zurich, ETH Zurich, 8093 Zurich, Switzerland}
\author{I. Cucchi}
\affiliation{Department of Quantum Matter Physics, University of Geneva, 24 Quai Ernest-Ansermet, 1211 Geneva 4, Switzerland}
\author{F. Y. Bruno}
\affiliation{Department of Quantum Matter Physics, University of Geneva, 24 Quai Ernest-Ansermet, 1211 Geneva 4, Switzerland}
\author{S. Ricc\`o}
\affiliation{Department of Quantum Matter Physics, University of Geneva, 24 Quai Ernest-Ansermet, 1211 Geneva 4, Switzerland}
\author{T.K. Kim}
\affiliation{Diamond Light Source, Harwell Campus, Didcot, United Kingdom}
\author{M. Hoesch}
\affiliation{Diamond Light Source, Harwell Campus, Didcot, United Kingdom}
\author{C. Barreteau}
\affiliation{Department of Quantum Matter Physics, University of Geneva, 24 Quai Ernest-Ansermet, 1211 Geneva 4, Switzerland}
\author{E. Giannini}
\affiliation{Department of Quantum Matter Physics, University of Geneva, 24 Quai Ernest-Ansermet, 1211 Geneva 4, Switzerland}
\author{C. Besnard}
\affiliation{Laboratoire de cristallographie, Ecole de Physique, University of Geneva, 24 Quai Ernest-Ansermet, 1211 Geneva 4, Switzerland}
\author{A. A. Soluyanov}
\affiliation{Theoretical Physics and Station Q Zurich, ETH Zurich, 8093 Zurich, Switzerland}
\affiliation{Department of Physics, St. Petersburg State University, St. Petersburg, 199034 Russia}
\author{F. Baumberger}
\affiliation{Department of Quantum Matter Physics, University of Geneva, 24 Quai Ernest-Ansermet, 1211 Geneva 4, Switzerland}
\affiliation{Swiss Light Source, Paul Scherrer Institut, CH-5232 Villigen PSI, Switzerland}

\date{\today}

\begin{abstract}
{We report a combined experimental and theoretical study of the candidate type-II Weyl semimetal \mote. Using laser-based angle-resolved photoemission we resolve multiple distinct Fermi arcs on the inequivalent top and bottom (001) surfaces. 
All surface states observed experimentally are reproduced by an electronic structure calculation for the experimental crystal structure that predicts a topological Weyl semimetal state with 8 type-II Weyl points. We further use systematic electronic structure calculations simulating different Weyl point arrangements to discuss the robustness of the identified Weyl semimetal state and the
topological character of Fermi arcs in \mote{}.
}
\end{abstract}

\maketitle

\section{Introduction}
The semimetallic transition metal dichalcogenides (W/Mo)Te$_2$ receive much attention for their unusual bulk electronic properties, including a non saturating magnetoresistance with values among the highest ever reported \cite{Ali2014} and pressure-induced superconductivity \cite{Pan2015a, Kang2015, Qi2015}. Very recent theoretical studies further predicted a new topological state of matter, dubbed type-II Weyl semimetal, in this series of compounds~\cite{Soluyanov2015a,Sun2015,Wang2015,Chang2016}. 
Weyl fermions in condensed matter arise as low energy excitations at topologically protected crossing points (Weyl points) between electron and hole bands~\cite{Wan2011, Weng2015a}. Weyl points always occur in pairs of opposite chirality and their existence near the Fermi level produces unique physical properties, including different magnetotransport anomalies~\cite{Nielsen83, Son2013, Hosur2013, Soluyanov2015a, Spivak2016, Burkov2015} and open Fermi arcs on the surface~\cite{Wan2011}, which were not observed in early angle resolved photoemission (ARPES) studies~\cite{Pletikosic2014,Jiang2015,Wu2015}. 

Weyl semimetals have been classified in type-I that exhibit point-like Fermi surfaces and respect Lorentz invariance, and type-II with strongly tilted Weyl cones appearing at the boundaries between electron and hole Fermi pockets and low-energy excitations breaking Lorentz invariance~\cite{Soluyanov2015a}.
Type-II Weyl semimetals necessarily have extended bulk Fermi surfaces. This renders the identification of the topological character of the surface Fermi arcs challenging since the Weyl points generally lie within the surface projected bulk states causing strong hybridization of bulk and surface states where the latter overlap with the bulk continuum. 
 
Experimentally, the realization of type-I Weyl fermions has been demonstrated in the TaAs family by angle-resolved photoemission (ARPES) and scanning tunneling spectroscopy~\cite{Xu2015a, Lv2015a, Lv2015, Yang2015, Xu2015,Inoue2016}, whereas the quest for type-II Weyls remains open.
Type-II Weyl fermions have first been predicted in \wte{}~\cite{Soluyanov2015a} crystallizing in the orthorhombic 1T' structure with broken inversion symmetry ($Pmn2_1$ space group). According to this study, \wte{} hosts 8 Weyl points $\sim 50$~meV above the Fermi level. However, the distance between Weyl points of opposite chirality is only 0.7\% of the reciprocal lattice vector rendering the observation of Fermi arcs connecting these points challenging for current spectroscopic techniques. Subsequently, it has been proposed that the low-temperature 1T' phase of \mote{} is a more robust type-II Weyl semimetal with topological Fermi arcs that are more extended in $k$-space~\cite{Sun2015, Wang2015}.

These predictions triggered a number of very recent ARPES studies reporting Fermi arcs on Mo$_{x}$We$_{1-x}$Te$_2$~\cite{Belopolski2015}, \mote{}~\cite{Huang2016arXiv,Deng2016arXiv,Zhou2016arXiv,Shi2016arXiv,Chen2016arXiv} and \wte~\cite{Bruno2016,Zhou2016arXivWTe2,Wu2016arXiv}. However, these studies do not agree on the bulk or surface character of the experimental Fermi surfaces, and provide conflicting
interpretations of the topological character of the surface states. The latter can in part be attributed to the theoretically predicted sensitivity of the number and arrangement of Weyl points in the Brillouin zone to the details of the crystal structure. Two such experimentally measured structures were discussed for \mote~\cite{Sun2015, Wang2015}. While the crystal structure used in the work of Refs.~\cite{Sun2015,Chen2016arXiv} was predicted to have 8 type-II Weyl points, appearing in the $k_z=0$ plane~\cite{Sun2015}, accompanied by 16 off-plane Weyl points~\cite{Wang2015}, the structure reported in the work of Ref.~\cite{Wang2015} was predicted to host only 4 type-II Weyl points formed by the valence and conduction bands. The differences in the Weyl point arrangements lead to distinct topological characters of Fermi arcs with similar dispersion.
These crystal structure intricacies, as well as the presence of two inequivalent surfaces in the non-inversion symmetric structure of \mote{} were largely ignored in recent works~\cite{Belopolski2015,Huang2016arXiv,Deng2016arXiv,Zhou2016arXiv,Shi2016arXiv,Chen2016arXiv}.

Here we use laser-based ARPES to clearly resolve distinct arc-like surface states on the two inequivalent (001) surfaces of \mote. To understand the topological nature of these arcs and their connection to type-II Weyl points, we present systematic calculations of the surface density of states simulating different Weyl-point arrangements. 
We find that the ARPES data from both surfaces is consistent with a topological Weyl semimetal state with 8 type-II Weyl points but does not conclusively establish this particular  number of Weyl points. Specifically,
we show that a large Fermi arc present on both surfaces and reported previously in Refs.~\cite{Huang2016arXiv,Deng2016arXiv,Zhou2016arXiv,Shi2016arXiv,Chen2016arXiv} is topologically trivial in our calculations and even persists for bulk band structures without Weyl points. 
On one of the surfaces we find additional small Fermi arcs extending out of the hole pockets. These short arcs are strong candidates for the topological surface states.
However, they resemble observations of Ref.~\cite{Chen2016arXiv} for a different Weyl point arrangement and 
 we argue that they are not robust in the sense that their emergence out of the bulk continuum is not topologically protected.

\section{Methods}
Single crystals of \mote{} in the monoclinic $\beta$-phase were grown by an optimized chemical vapor transport method (see Appendix). The low temperature 1T' crystal structure was characterized using single-crystal X-ray diffraction (Rigaku Supernova diffractometer, Mo~K$\alpha$ radiation, Oxford Instrument cryojet cooling system).  Synchrotron based ARPES experiments were performed at the I05 beamline of Diamond Light Source using photon energies of $40-90$~eV. Laser-ARPES experiments were performed with a frequency converted diode-laser (LEOS solutions) providing continuous-wave radiation with 206~nm wavelength ($h\nu = 6.01$~eV) focused in a spot of $\sim 5$~$\mu$m diameter on the sample surface and an MBS electron spectrometer permitting two dimensional  $k$-space scans without rotating the sample. Samples were cleaved \textit{in-situ} along the ab-plane at temperatures $< 20$~K. Measurement temperatures ranged from 6~K to 20~K and the energy and momentum resolutions were $\sim 15$~meV/0.02~\AA$^{-1}$ and 2~meV/0.003~\AA$^{-1}$ for synchrotron and laser ARPES experiments, respectively. Electronic structure calculations were done using the VASP software package~\cite{Kresse96} with PAW pseudopotentials~\cite{Blochl94} that include spin-orbit coupling. The PBE functional~\cite{Perdew96} was used in the exchange-correlation potential. A 16$\times$10$\times$4 $\Gamma$-centered $k$-point mesh was used for Brillouin zone sampling, and the energy cutoff was set to 450~eV. The Wannier-based projected tight-binding models~\cite{Marzari1997, Souza2001, Mostofi2014} capturing all the $d$-states of Mo and $p$-states of Te were used to analyze the surface density of states.
Surface spectra were calculated by the software package Wannier\_tools~\cite{wannier_tools}, which is based on the iterative Green's function~\cite{Sancho1985}.

\begin{figure}[!tb]
\includegraphics[width=0.47\textwidth]{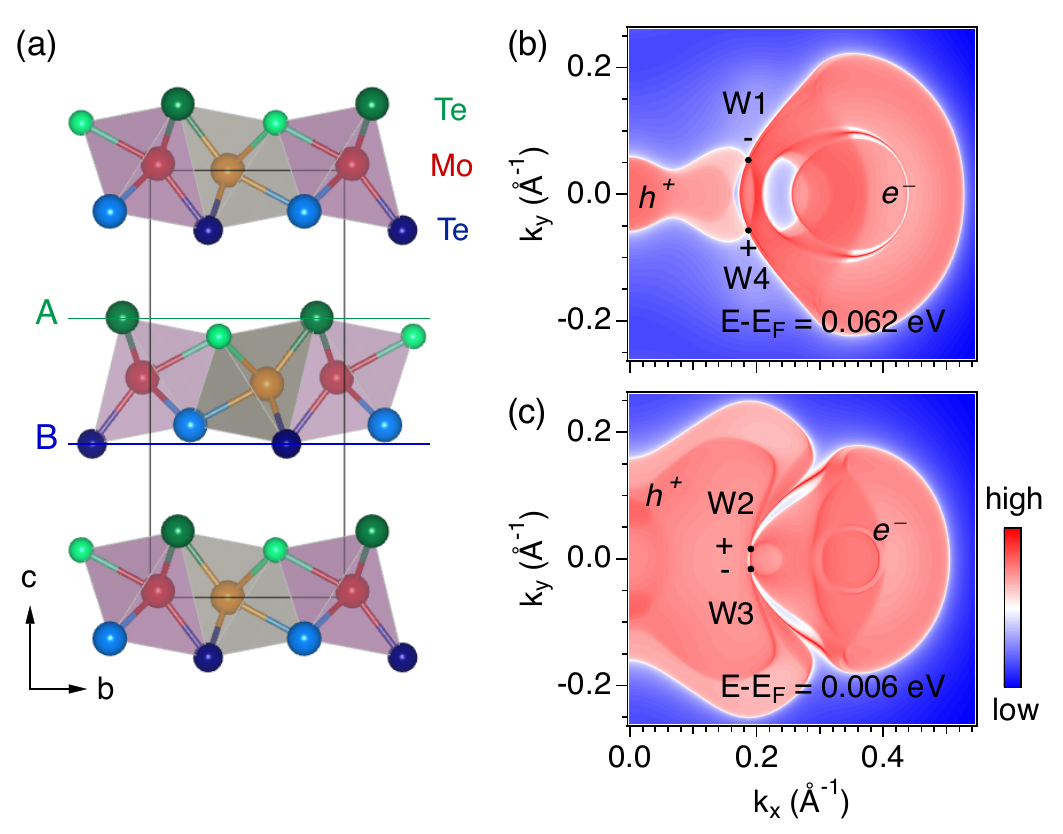} %
\caption{Crystal structure and Weyl points of \mote.  (a) Crystal structure of the 1T' phase (space group \textit{Pmn}2$_{1}$) determined by X-ray diffraction. The two inequivalent surfaces, labelled $A$ and $B$, are indicated. 
(b,c) Momentum resolved density of states of the (001)-surface at $E - E_{F} = 0.006$~eV and 0.062~eV, respectively, illustrating the touching points W1,4 and W2,3 of electron and hole pockets. W1(W2) is a mirror image of W4(W3).
}
\label{f1}
\end{figure}

\section{Results}
The monoclinic $\beta$-phase of \mote{} undergoes a structural phase transition to the orthorhombic 1T' phase (also called $\gamma$ or T$_d$ phase) at $\sim 250$~K~\cite{Clarke1978}. The low-temperature 1T' phase shares the non-centrosymmetric \textit{Pmn}2$_{1}$ space group with \wte{} and consists of double layers of buckled tellurium atoms bound together by interleaving molybdenum atoms. The resulting \mote{} layers are stacked along the $c$ axis with van der Waals interlayer bonding as illustrated in \one(a).
Note that the broken inversion symmetry of the 1T' structure implies the existence of two inequivalent (001) surfaces with the [100] axis pointing out or into the surface, respectively. We denote these surfaces by $A$ and $B$, as indicated in \one(a).
In \one (b-c) we illustrate the number and positions of Weyl points in \mote{}, calculated for the experimental crystal structure of our samples with lattice constants $a = 3.468$~\AA, $b = 6.310$~\AA, $c = 13.861$~\AA{} determined at $\rm{T}=100$~K (see Figs.~\ref{fs1} and ~\ref{fs2} in the Appendix for the temperature dependence of the lattice constants and calculations with unit cell parameters extrapolated to different temperatures).

\begin{figure*}[tb]
\includegraphics[width=0.95\textwidth]{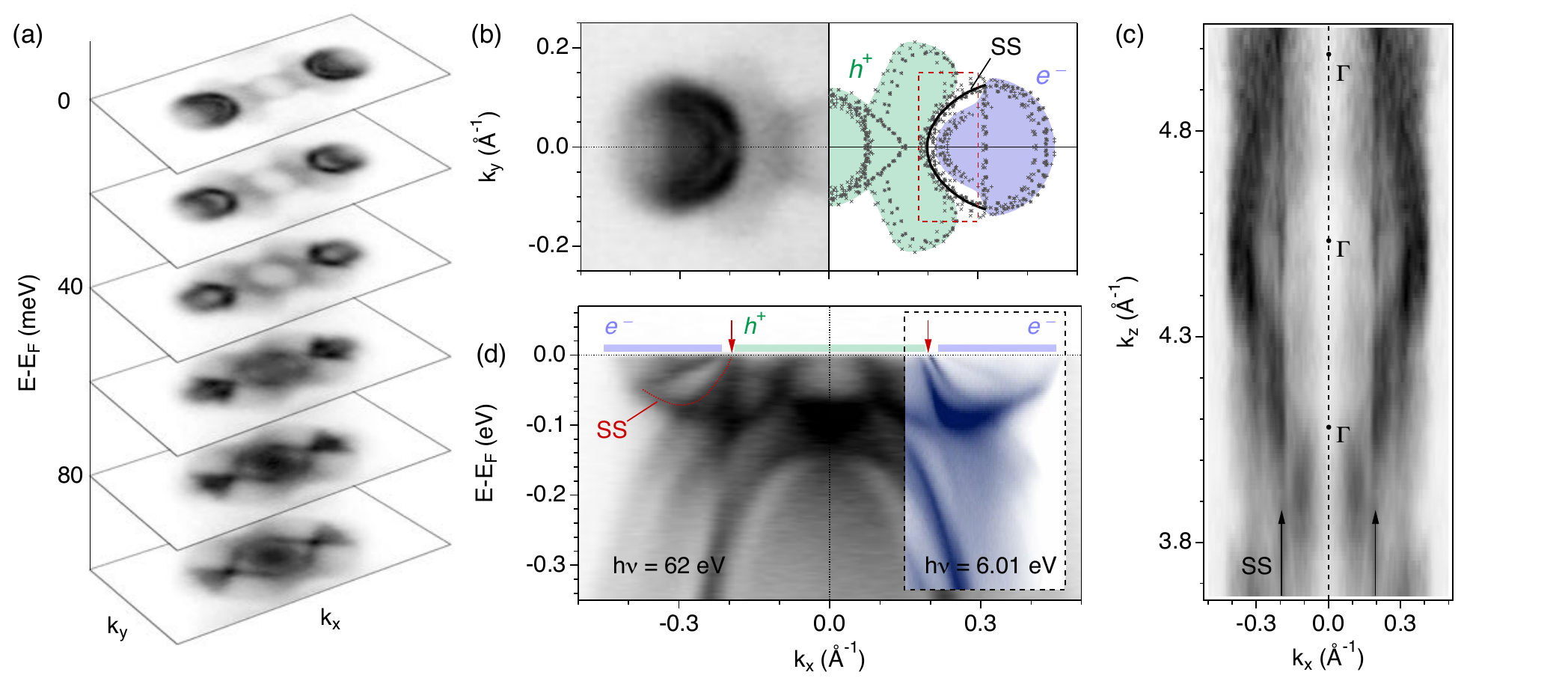} %
\caption{Electronic structure of  \mote. (a) Stack of constant energy maps measured with 62~eV photon energy and $p$-polarization. (b) Left: ARPES Fermi surface at 62~eV photon energy. Right: Fermi surface contours extracted from Fermi surface maps measured at different photon energies. The largest extension of electron and hole pockets corresponding to the projected bulk band structure is indicated in blue and green respectively. The thin black arc between hole and electron pockets is a surface state. (c) $k_{z}$ dependence of the electronic states at the Fermi level in the k$_{y} = 0$ plane. The straight lines at $k_{x}=\pm$ 0.195 \AA$^{-1}$ marked by black arrows are the signature of a two-dimensional electronic state at the surface of \mote. Values of $k_{z}$ have been calculated for free electron final states with an inner potential of 13~eV. (d) Overview of the surface band structure measured along the high symmetry direction k$_{y} = 0$ taken with $h\nu=62$~eV and 6.01~eV (blue inset). }
\label{f2}
\end{figure*}

Unlike Ref.~\cite{Wang2015} that predicted 4 Weyl points for a structure with 0.3\% smaller lattice constant $a$, our new calculation finds 8 type-II Weyl points between the valence and conduction band in the  k$_{z} = 0$ plane. We checked that no further off-plane Weyl points
are present in our calculation. Note, however, that additional Weyl points  above the Fermi level formed solely by the conduction bands were found in Ref.~\cite{Wang2015} (some of them in the $k_z=0$ plane).
Given that the gaps separating different bands are small in \mote, even weak external perturbations may cause changes in band ordering, resulting in the appearance of additional Weyl points.   
This illustrates the exceptional sensitivity of key topological properties of \mote{} to minute changes in the structure as already pointed out in Refs.~\cite{Wang2015,Sun2015}.
This problem is aggravated by the variation of the lattice constants of \mote{} grown under different conditions, which might arise from a slight off-stoichiometry and/or the proximity of the hexagonal 2H phase in the non linear temperature-composition phase diagram of \mote{} favoring inter-growth of different phases~\cite{Brewer:1980aa}.

In \two{} we present the overall band structure of \mote{} determined by ARPES. The Fermi surface consists of a large hole-like sheet centered at the $\Gamma$ point and two symmetric electron-like Fermi surfaces, one at positive and one at negative $k_x$ values, that nearly touch the hole pocket. The character of the different Fermi surfaces can be inferred from the stack of constant energy maps in \two(a) illustrating how the sizes of electron and hole pockets decrease and increase with energy, respectively. The dominant spectral feature at the Fermi surface is a symmetric arc-like contour in the narrow gap between electron and hole pockets (black line in \two(b)), which is not seen in bulk band structure calculations. The $k_x-k_z$ Fermi surface map in \two(c) shows that this state (SS) does not disperse in $k_{z}$ over the entire Brillouin zone as expected for a strictly two-dimensional electronic state localized at the surface. \two(d) shows the overall band dispersion along the high symmetry direction $k_{y}=0$. The surface state dispersing from the bottom of the electron bands up to the hole bands is indicated by a thin red line. The same cut measured with high-resolution laser ARPES (blue inset) reveals a more complex situation. Rather than a single surface state we can resolve two very sharp dispersing states with similar Fermi velocities. While the outer one at larger $k_x$ values shows a clear Fermi level crossing, the inner one, which is most intense $\sim100$~meV below E$_{F}$, looses most of its spectral weight approaching E$_{F}$.

\begin{figure*}[t]
\includegraphics[width=0.75\textwidth]{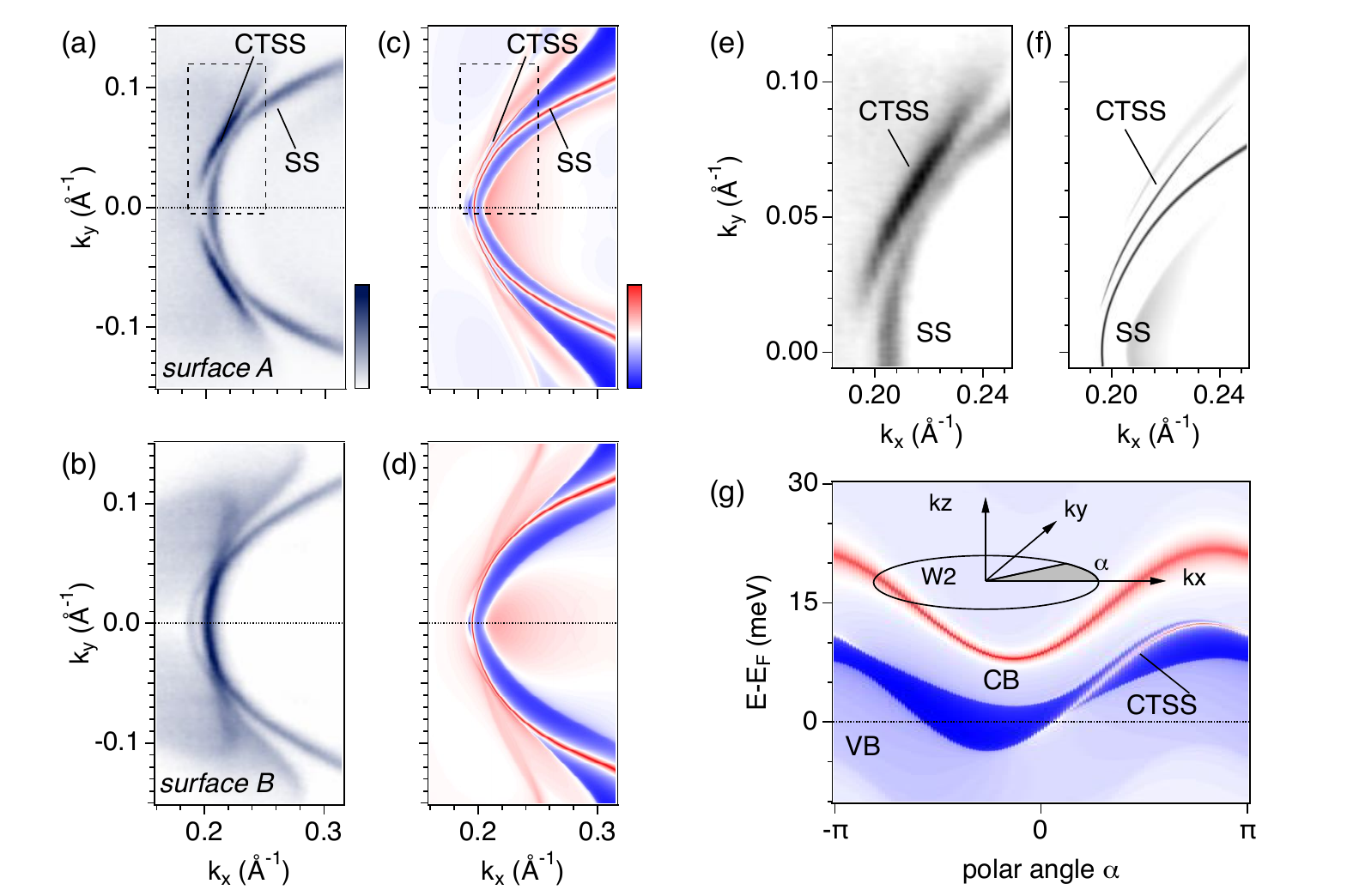} %
\caption{Fermi arcs on the surface of \mote. (a,b) Detailed view of the Fermi surface measured with 6.01~eV photon energy and $p$-polarization on the $A$ and $B$ surface terminations of \mote. The photoemission intensity has been symmetrized with respect to the  $k_{x}$ axis. (c,d) Surface density of states for the two surface terminations calculated 10~meV below the Fermi level. (e,f) Zoom into the areas enclosed by a dashed rectangle in (a) and (b), respectively. (g) Numerical results for the surface density of states along an elliptical contour (with $r_{kx}=0.0018$~\AA$^{-1}$, $r_{ky}=0.005$~\AA$^{-1}$) enclosing the Weyl point W2. The topological surface state is connecting valence to conduction bands and produces CTSS in the calculations.
}
\label{f3}
\end{figure*}

In \three, we zoom into the region enclosed by a red dashed rectangle in \two(b). Using a laser focused into a spot of $\sim5$~$\mu$m diameter on the sample surface as excitation source and high energy and momentum resolution we are able to identify two clearly distinct Fermi surfaces appearing with the same abundance on a large number of cleaved samples.
For surface $A$ we find two sets of intense and  sharp contours, a large arc-like state (SS) and two shorter arcs with weak curvature that are visible in a narrow range of momenta $|k_{y}|\lesssim0.08$ \AA$^{-1}$ and vanish again in the $k_{y} = 0$ plane (\three(a)). On the $B$-type surface, we resolve two large arcs with different curvatures separated by a small non-crossing gap (\three(b)).
These signatures are well reproduced by calculations of the surface density of states for the inequivalent top and bottom surface of \mote{} shown in \three(c,d).

According to the calculations presented in \three{} and in the Appendix, the large Fermi arc (SS) is topologically trivial for the present crystal structure, and even persists in calculations for different structures that exhibit no bulk Weyl points at all.
The shorter arc  (CTSS for candidate topological surface state), on the other hand, is a part of the surface state that connects valence to conduction states, as shown for the elliptical path surrounding the projection of the W2 point onto the surface (\three(g)). This elliptical path can be viewed as a cut of a cylinder that encloses the W2 point, and the axis of which is aligned along $[001]$. The valence and conduction bands are gapped at all $k$-points on this cylinder, and thus a Chern number can be computed for the occupied states, corresponding to the chirality of the enclosed Weyl point. We find this number to be $+1$, and hence CTSS in \three(g) is topologically protected. The calculation agrees remarkably well with the experiment (\three(e,f)). Therefore, CTSS is a strong candidate for a topological surface state.

\begin{figure*}[t]
\includegraphics[width=0.7\textwidth]{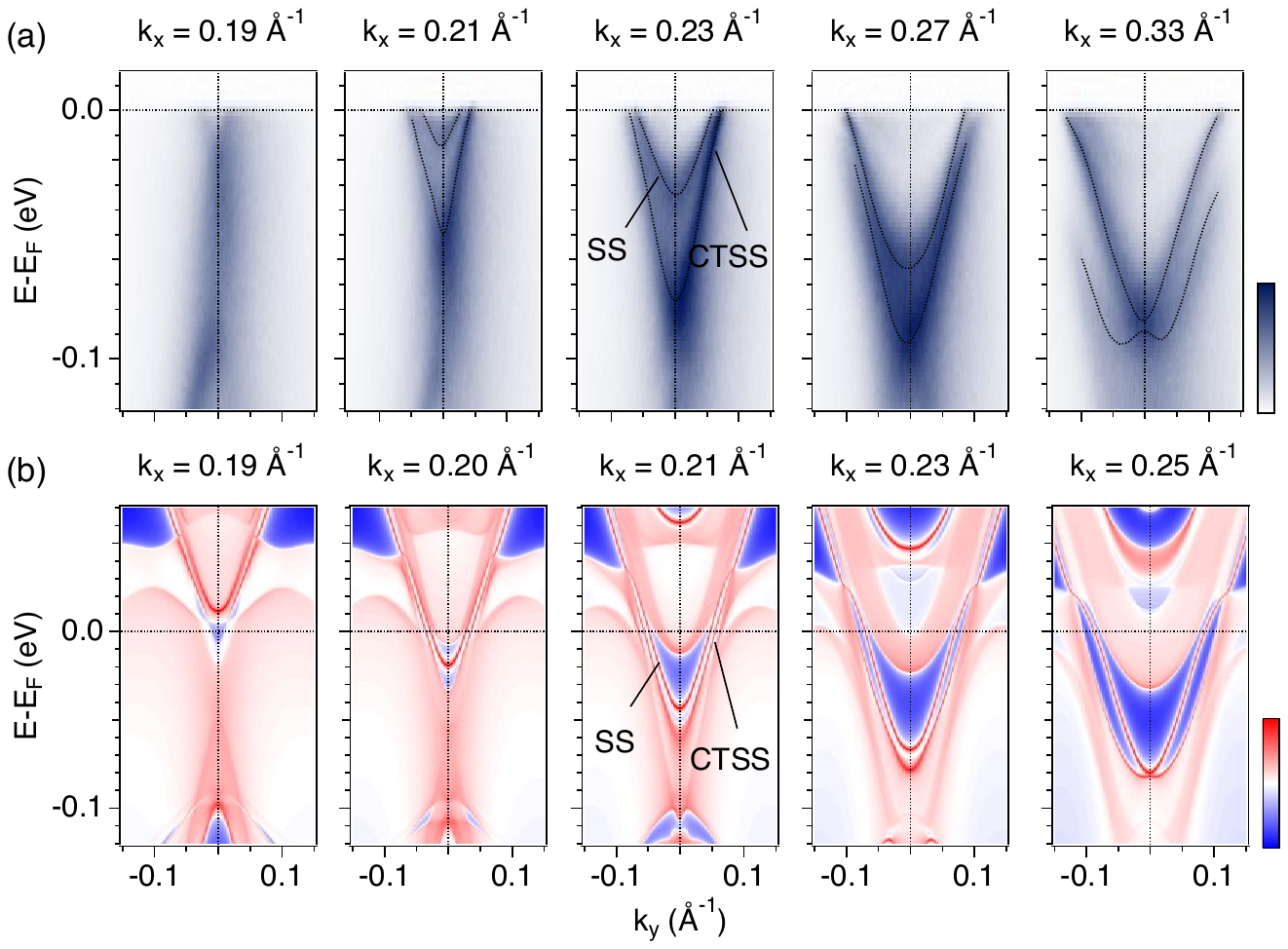} %
\caption{(a) Dispersion plots taken on surface $A$ parallel to $k_{y}$ for different values of $k_{x}$. (b) Calculated surface energy dispersion along equivalent cuts in momentum space for the same surface termination.}
\label{f4}
\end{figure*}
The good agreement between ARPES data and numerical calculations extends to the dispersion plots shown in \four(a,b) taken parallel to the $k_{y}$ axis for different values of $k_{x}$. The trivial surface state SS has a nearly parabolic dispersion in our experiments and well defined Fermi crossings for $k_{x}\geq 0.21$~\AA$^{-1}$, whereas CTSS has a more complex evolution of the dispersion and only reaches the Fermi level over a very limited range of momenta (second and third panel of \four(a)).
This behavior is well reproduced by the calculated surface density of states (\four(b)) showing that CTSS is only defined for energies and momenta where it emerges out of the projected bulk hole pocket.
For this reason it is not possible to derive the positions of type-II Weyl points from the extensions of the topological arcs at the Fermi surface.

We note that the intensity of the bulk bands is strongly suppressed in our laser-ARPES data. This is common in very low-energy photoemission and arises from the band-like character of final states leading to strongly photon energy dependent matrix elements of bulk states. We further find that all bulk states are broad compared to SS and CTSS. This can be explained naturally by the surface sensitivity of photoemission causing an intrinsically poor $k_z$ resolution. For the large unit cell of \mote, we estimate that $\Delta k_z > 0.2\pi/c$, which implies a substantial broadening of states that disperse in $k_z$. The absence of such broadening in CTSS indicates that this state is highly two-dimensional consistent with its identification as surface state.

CTSS found in the calculations connects Weyl points that both project into the bulk hole pocket at the Fermi level. Moreover, all the Fermi pockets  in \mote{} have zero Chern numbers. For these reasons the emergence of Fermi arcs from the hole pocket is not dictated by topology, but is rather a consequence of a particular realization of the shape of CTSS. In general, a topologically trivial surface state can also emerge out of the topologically trivial pocket, as is the case at hand for SS.
The intricate relation of CTSS to the Weyl points of the system is evident from the fact that a Fermi arc with very similar dispersion
was also reported in the calculations of Refs.~\cite{Sun2015,Chen2016arXiv} 
with different in-plane Weyl point positions and 16 additional off-plane Weyl points~\cite{Wang2015}.
In that case additional surface resonances or surface states hybridized with each other can appear inside or outside the nearby bulk continuum due to the presence of the off-plane Weyl points. 
In principle, such resonances may cause the observed state to be topologically trivial in analogy to SS. On the other hand, CTSS in our calculations and its analogue in Refs.~\cite{Sun2015,Chen2016arXiv} are both topological
and we do not find surface states with similar shape that are topologically trivial in any of our calculations.
The question is therefore whether the absence of a topologically trivial scenario at hand can be taken as positive proof of a Weyl semimetal state in \mote. The systematic nature of calculations carried out by us and others suggest that this is reasonable even though evidence of absence (in this case of a topologically trivial scenario explaining CTSS) is never exhaustive.

CTSS is only observed clearly on one of the two inequivalent (001) surfaces. This can be explained by noticing that the trivial value of the $\mathbb{Z}_{2}$ topological invariant~\cite{Kane2005} for the $k_y=0$ plane in the Brillouin zone of MoTe$_2$ is consistent with two possible arc connectivities: the one consistent with CTSS, assuming that a surface state connects the projections of W1 and W2 points, and the other, in which the arcs connect the projections of W1,2 with their mirror images W3,4. While the ARPES data of the $B$-type surface (\three(b)) is consistent with the latter scenario, 
 they do not uniquely define a particular Weyl point arrangement and connectivity.

To illustrate this point we show in \five(c,d) the surface density of states on the $B$ type surface for crystal structures resulting in 4~\cite{Wang2015} and 8 (this work) type-II Weyl points.
Both calculations clearly show a large and fully spin-polarized Fermi arc with virtually identical dispersion. However, in the structure of Ref.~\cite{Wang2015} with 4 Weyl points (\five(c)), this arc is topological, while for the crystal structure of this work, we find that the arc is trivial (\five(d)). As shown in figures~\ref{fs5} and \ref{fs6} in the Appendix, the large Fermi arc even persist in calculations for \mote{} and \wte{} finding zero Weyl points. 
This demonstrate that the topological character of Fermi arcs in (Mo/W)Te$_2$ can in general not be uniquely deduced from a comparison of experimental and theoretical band dispersions. It further shows that neither the experimental observation of the large Fermi arc in \mote{} or \wte{} nor of its spin-structure is a suitable fingerprint to robustly identify the type-II Weyl state in this series of compounds.The observation of CTSS on the other hand cannot readily be explained in a topologically trivial scenario and thus does provide evidence for a type-II Weyl semimetal state.

\begin{figure}[tb]
\includegraphics[width=0.47\textwidth]{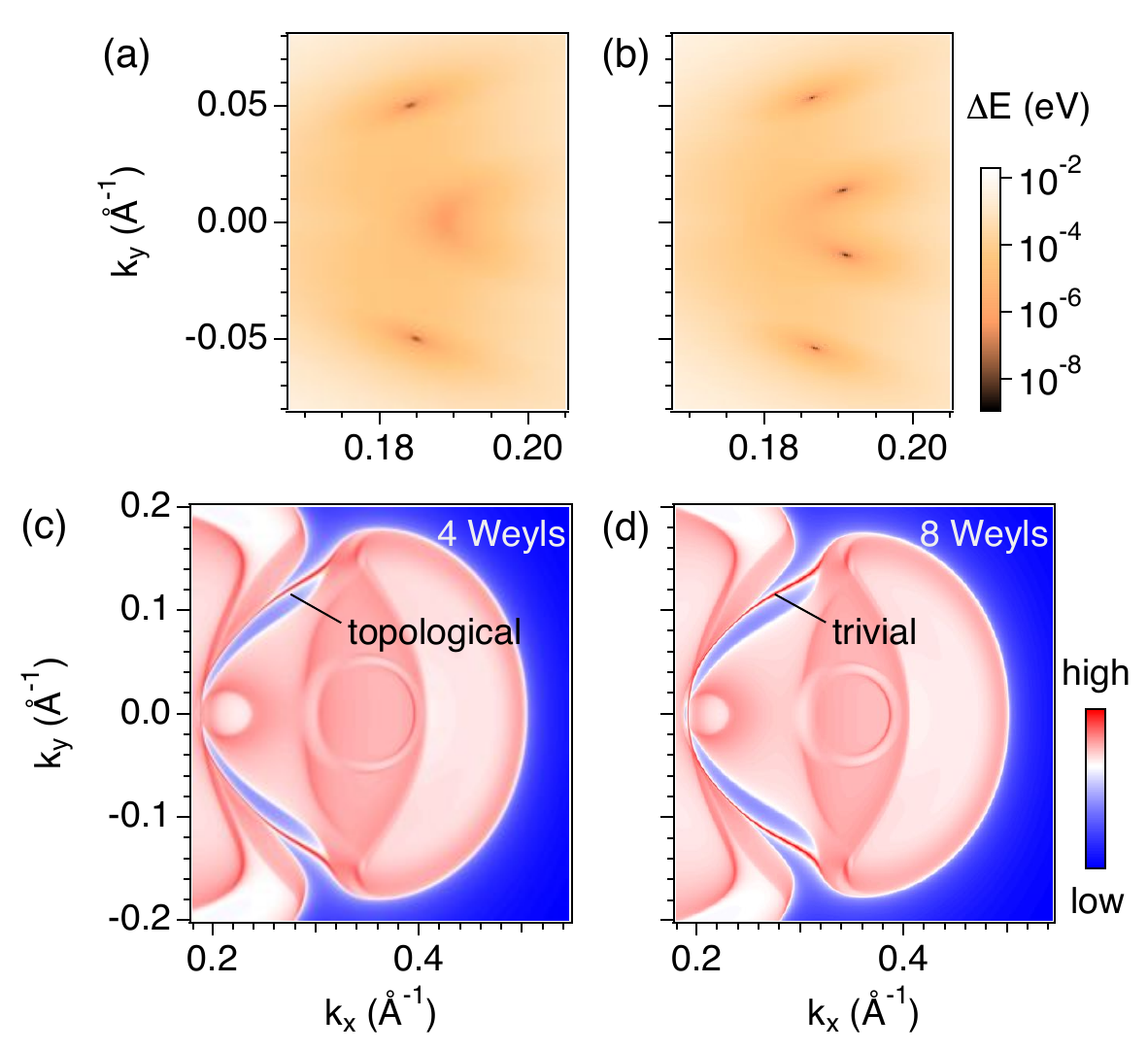} %
\caption{Trivial and topological Fermi arcs on the $B$-type surface of \mote. (a) Position of the 4 Weyl points calculated for the experimental crystal structure reported in Ref.~\cite{Wang2015}. (b) Using the structure determined in our work, we find 8 Weyl points. (c,d) Theoretical Fermi surfaces for the structures used in (a) and (b), respectively. The large Fermi arc has a virtually identical dispersion in the two scenarios but is topological in (c) and trivial in (d).
}
\label{f5}
\end{figure}

\section{Conclusions}
We conclude that \mote{} is a strong candidate for the realization of a type-II Weyl semimetal phase. 
All the experimental observations presented here agree with our numerical calculations finding 8 type-II Weyl points
and a coexistence of topological and trivial Fermi arcs.
While the data are also consistent with other possible Weyl point arrangements, systematic calculations do not show a topologically trivial scenario that could explain all the experimental observations.
We point out that an ideal candidate type-II Weyl semimetal permitting a more direct experimental identification should possess projected electron and hole pockets with non-zero total Chern number, thus guaranteeing the appearance of the surface state in both numerical and experimental data.

\begin{acknowledgements}
We gratefully acknowledge discussions with G. Aut\'{e}s, A. Bernevig, R. Cerny, A. Morpurgo, C. Renner. This work was supported by the Swiss National Science Foundation (200021-146995, 200021-153405, PZ00P2-161327). Q.S.W. and A.A.S. acknowledge funding from Microsoft Research and the Swiss National Science Foundation through the National Competence Centers in Research MARVEL and QSIT. Q.S.W. was also supported by the National Natural Science Foundation of China (11404024). The authors acknowledge Diamond Light Source for time on beamline I05 under proposal SI12404.
\end{acknowledgements}

\section{Appendix}
\subsection{Crystal growth}
Single crystals of monoclinic $\beta$-MoTe$_{2}$ were grown by the chemical vapor transport technique using TeCl$_{4}$ as a transport agent. The pure elements Mo and Te and the transport agent TeCl$_{4}$ were mixed in a stoichiometric anion ratio, according to the reaction equation:  
\[2 \bigg(\frac{n-1}{n}\bigg)\textrm{Te} + \bigg(\frac{1}{n}\bigg)\textrm{TeCl}_4 + \textrm{Mo} \to \textrm{MoTe}_2 +\bigg(\frac{3}{n}\bigg)\textrm{Cl}_2\]
with $n=10$. The total weight of each sample was $\sim$ 0.2$\div$0.3 g. The mixture was weighted in a glovebox and sealed under vacuum ($5 \times 10^{-6}$~mbar) in a quartz ampule with an internal diameter of 8 mm and a length of 120 mm. The sealed quartz reactor was placed in a two zone furnace in the presence of a thermal gradient dT/dx $\approx 5^{\circ}\div10^{\circ}$~C/cm, and heated up to temperatures T$_{\rm{hot}}= 980^{\circ}$C at the hot end and T$_{\rm{cold}}= 900^{\circ}$C at the cold end. After 1 week of growth, the quartz ampule was quenched in air to yield the monoclinic phase, $\beta$-MoTe$_{2}$. The crystals obtained on the cold side were shiny-grey and rectangular. \\

\subsection{Lattice constants and Weyl points}

\begin{figure}[b]
\includegraphics[width=7.5cm]{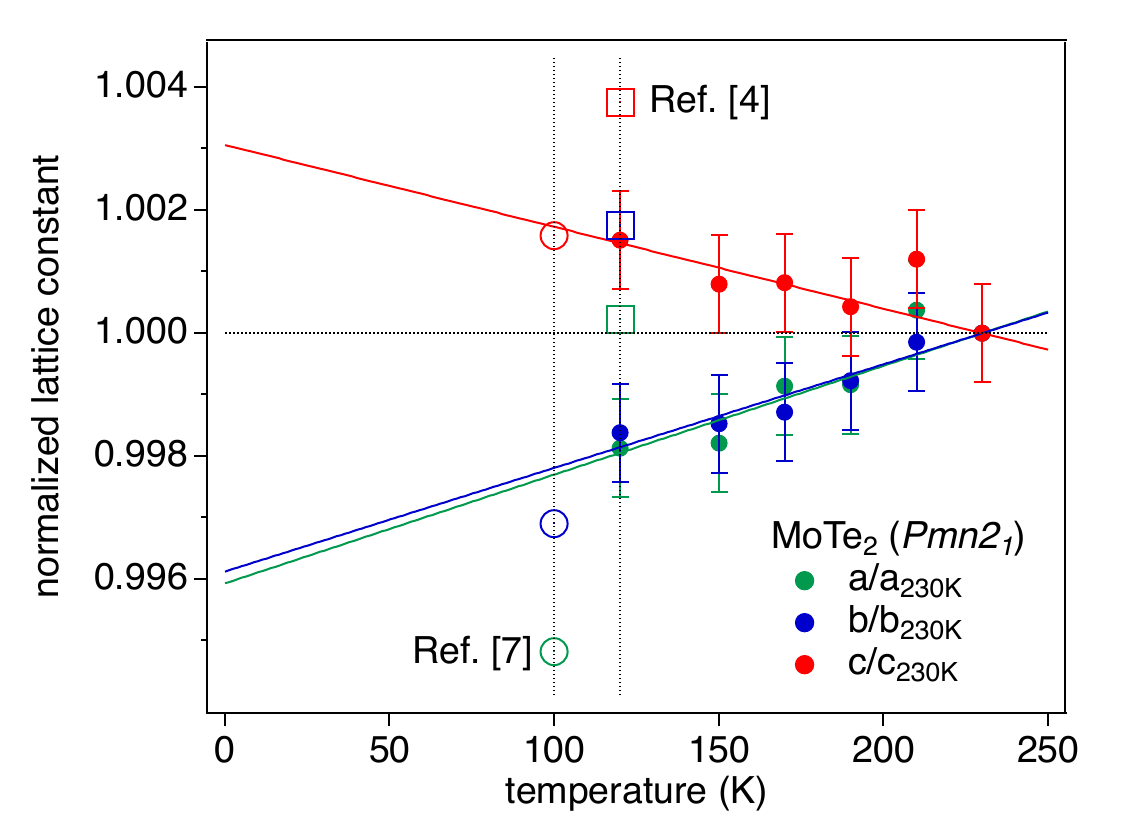} %
\caption{ Temperature dependence of the lattice constants of \mote{} in the 1T' phase (space group $Pmn2_{1}$). Open circles and squares are the lattice constants of \mote{} obtained with different sample growth procedures as reported in Refs.\cite{Wang2015, Qi2015}. All values have been normalized by the lattice parameters of our samples measured at 230 K ($a = 3.4762$~\AA, $b = 6.3239$~\AA, $c = 13.837$~\AA).}
\label{fs1}
\end{figure}

\begin{figure}[tb]
\includegraphics[width=8.3cm]{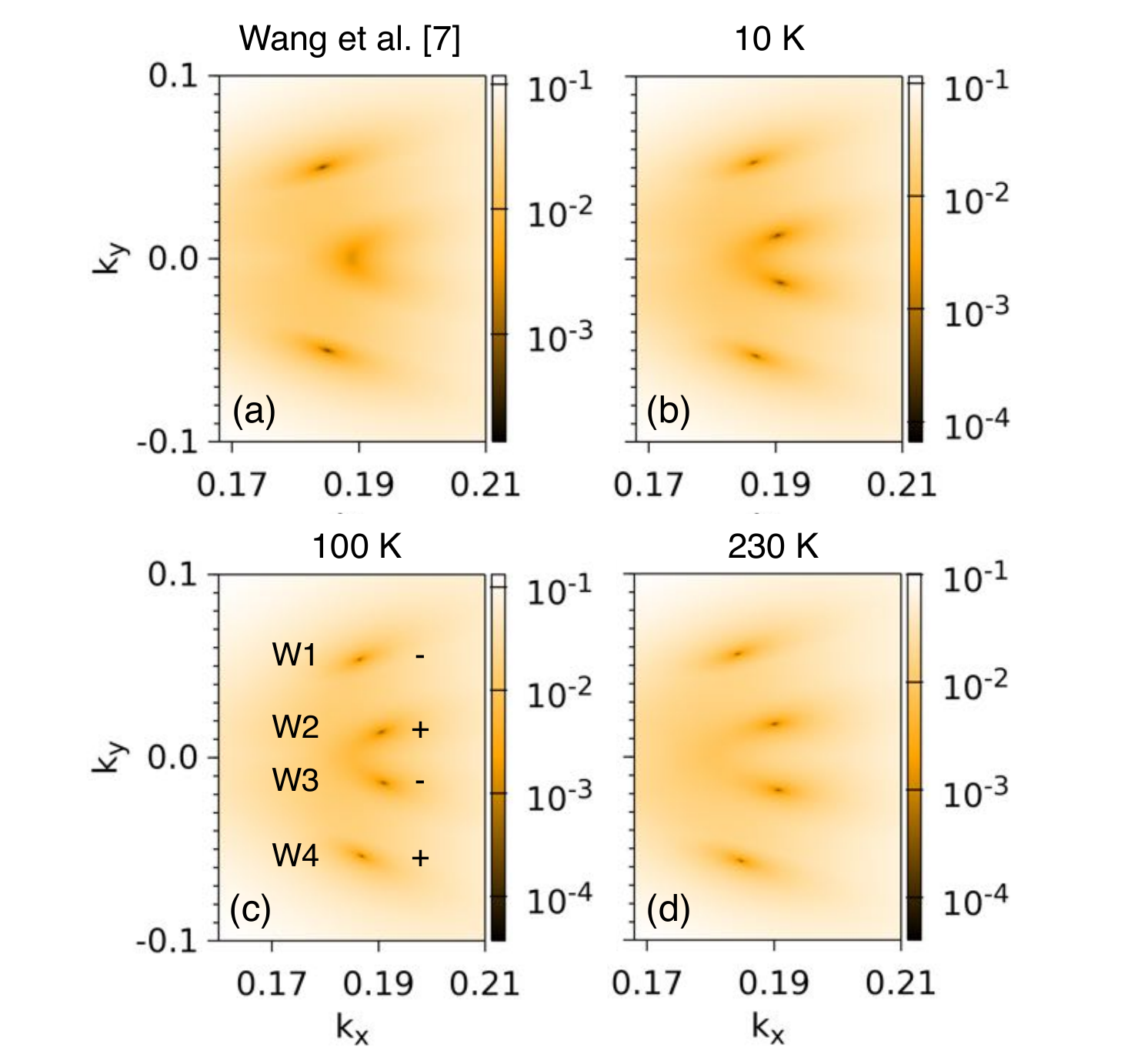} %
\caption{False color plot of the energy gap between valence and conduction bands in the $k_{z}$ = 0 plane illustrating how the number and position of Weyl points evolve for different unit cell parameters: (a) Structure in Ref.~\cite{Wang2015} which has a 0.3\% smaller lattice constant $a$; (b-d) Lattice constants at 10~K, 100~K and 230~K obtained from a linear extrapolation based on the experimental data in \ones.
The coordinates of the Weyl points found for the unit cell parameters at 100~K used in the main text are $(0.1855, \pm0.0539, 0)$~\AA$^{-1}$ and $E-E_F = 0.062$~eV for W1,4 and $(0.1906, \pm0.0152, 0)$~\AA$^{-1}$ and $E-E_F = 0.006$~eV for W2,3. The remaining 4 points are mirror symmetric with respect to $k_{x}$
}
\label{fs2}
\end{figure}

The monoclinic $\beta$-phase of \mote{} (unit cell parameters at 280~K of $a = 3.479$~\AA, $b = 6.332$~\AA, $c = 13.832$~\AA, $\gamma = 93.83^{\circ}$) undergoes a structural phase transition to the orthorhombic 1T' phase at $\sim 250$~K.
\ones{} displays the temperature dependence of the lattice constants in the 1T' phase of \mote. Consistent with the early study of Clarke \textit{et al.}~\cite{Clarke1978}, we find that the $c$ axis expands with decreasing temperature while $a$ and $b$ decrease slightly. For comparison we have also displayed the lattice constants of \mote{} from the two previously reported structures  in Refs.~\cite{Wang2015, Qi2015}. 

In the main text we have compared ARPES data measured below 20~K with calculations based on the lattice constants at 100~K extrapolated linearly from the experimental values measured between 230~K and 120~K.
This is a reasonable approach as we do not expect significant changes of the unit cell parameters between 100~K and the temperature of the ARPES experiments. The thermal expansion coefficient is usually negligible below $\sim 100$~K and no additional structural phase transitions have been reported for \mote{} at these low temperatures.
In order to confirm the validity of this comparison we performed additional calculations for different sets of lattice parameters assuming constant thermal expansion. As shown in \twos(b-d), calculations based on the lattice constants of our structure at 10~K, 100~K and 230~K estimated in this way find the same number of Weyl points at nearly identical positions. All these structures result in 8 Weyl points at $k_{z}$ =0 and no additional off-plane Weyls points, formed between valence and conduction bands. This indicates that the contraction of the lattice upon cooling will have only a minor effect in our samples and will not change the topological character of the Fermi arcs discussed in the main text. Note, however, that \twos{} reveals a clear trend in the position and number of Weyl points with the in-plane lattice constant $a$. As $a$ gets smaller, due to thermal contraction (b-d) or a different growth method (a), the Weyl points W2 and W3 move closer to the k$_{x}$ axis and will eventually merge and annihilate. \\

\begin{figure*}[]
\includegraphics[width=0.9\textwidth]{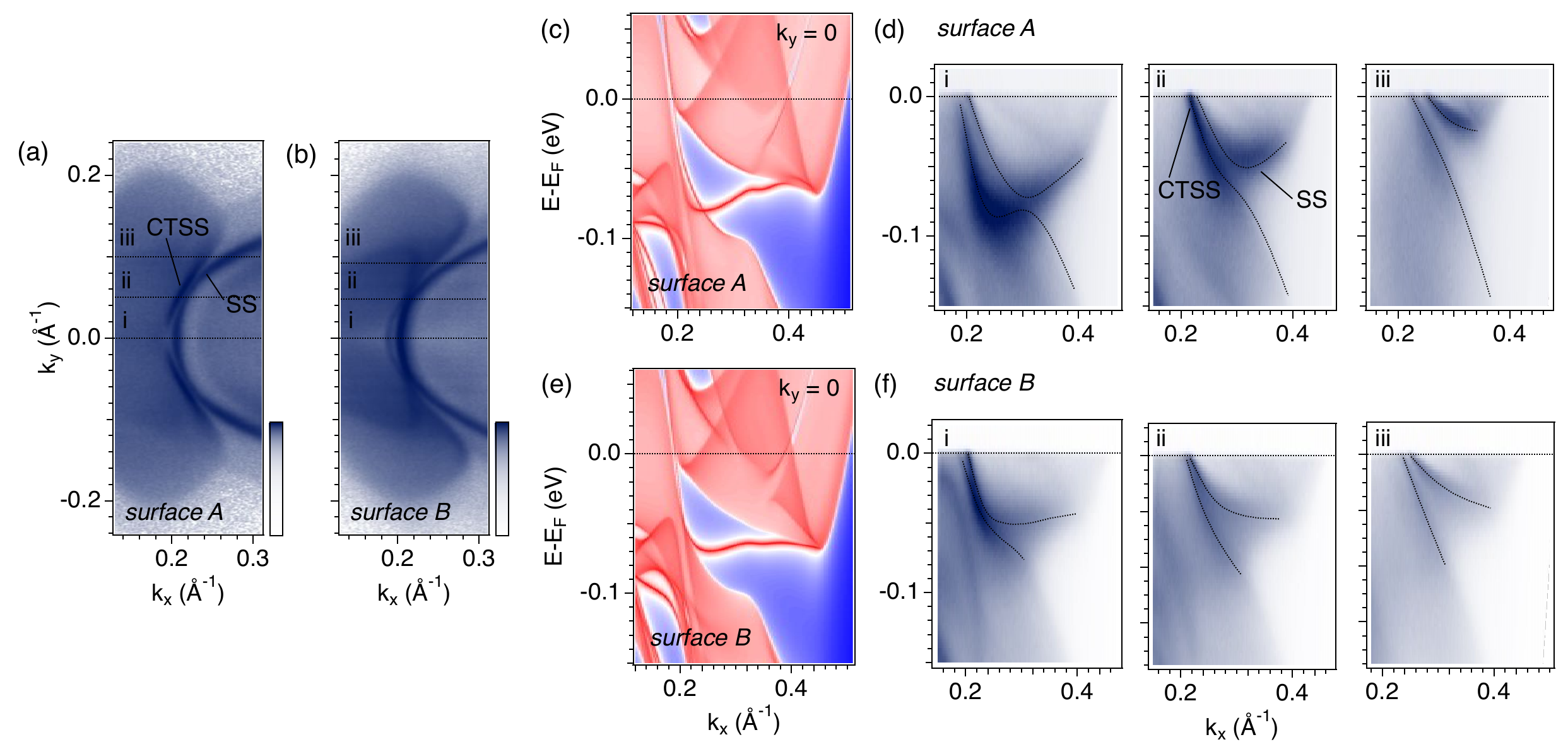} %
\caption{(a,b) Detailed view of the Fermi surface measured with 6.01~eV photon energy and $p$ polarization on the $A$ and $B$-type surfaces of \mote, displayed on a logarithmic color scale. 
This photon energy corresponds to $k_z$ values around $0.6\pi/c$ in the fifth Brillouin zone assuming a free electron final state with an inner potential of 13~eV. Note, however, that a free electron final state model is not fully appropriate at such low photon energies, which results in a substantial uncertainty of $k_z$ values.
(c,e) Surface energy dispersion along the $k_{y}=0$ direction calculated for the top and bottom surface, respectively. (d,f) Dispersion plots measured on surface $A$ and $B$ parallel to the $k_{y}=0$ direction at the $k_{y}$ values ``i" to ``iii" marked in panels (a) and (b), respectively.} 
\label{fs3}
\end{figure*}

\subsection{Surface state dispersion parallel to the $k_{y} = 0$ direction}

In \threes{} we present dispersion plots measured parallel to the $k_{y}$=0 direction on the two different surfaces of \mote. In each panel of \threes(d,f) we can identify two dispersing states that contribute to the trivial arc (SS) and the candidate topological arc (CTSS) at the Fermi surface. 
Their dispersion is clearly distinct on the two surface types. In particular, on the $A$-type surface, SS is more dispersive near the bottom of the bulk electron pocket and the two surface states cross each other along the high symmetry cut with $k_{y} = 0$, while this is not observed on the $B$-type surface. These differences are well reproduced by calculations of the surface band structure for a semi-infinite crystal (panels (c,e)) allowing us to unambiguously identify surface $A$ and $B$ as the top and bottom (001) surface of \mote. 

\threes{} also provides further evidence for the surface character of CTSS. In particular, the clearly distinct dispersion of this spectral feature below the Fermi level is hard to reconcile with a bulk origin, since bulk states are not expected to differ between the two surface types.
Additionally, we note that all bulk states are substantially broader and much less intense than surface related states throughout the entire Brillouin zone. This is evident from panels (a,b) where we use a logarithmic color scale to enhance the projected bulk Fermi surfaces.

\begin{figure*}[]
\includegraphics[width=16cm]{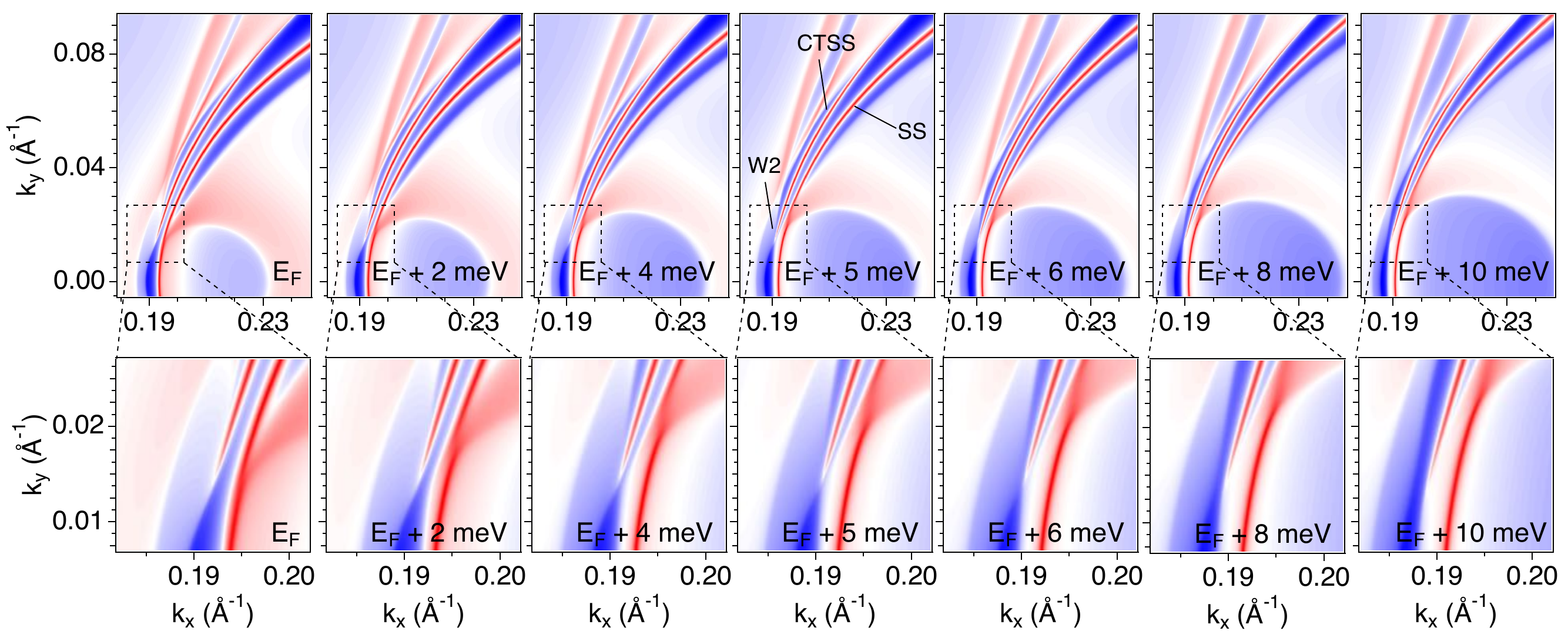} %
\caption{Momentum-resolved surface density of states illustrating the evolution of the topological Fermi arc (CTSS) around the energy of the W2 Weyl point.}
\label{fs4}
\end{figure*}

\subsection{Candidate topological Fermi arc}

In Fig.~3(g) of the main text, we show that the candidate topological surface state (CTSS) identified in our ARPES data connects the valence (hole) and conduction (electron) band along a fully gapped path enclosing a single Weyl point with Chern number $+1$ (W2) and is thus topologically non-trivial. In \fours{} we use more conventional $k$-space maps at different, finely-spaced energies to illustrate how this state connects the valence and conduction bands. At the Fermi level, a small gap separates the momentum resolved surface density of states of the electron and hole pockets and the CTSS arc emerges on both ends out of the hole pocket. As the energy is increased towards the Weyl point W2 at $+6$~meV, this gap reduces and the electron and hole pockets eventually touch. At higher energy, the CTSS arc emerges out of the electron pocket at $k_{y}\approx0.015$~\AA$^{-1}$ and disappears into the hole pocket at $k_{y}\approx0.07$~\AA$^{-1}$.
 
\subsection{Topological character of the large Fermi arc}

We have shown in the main text that the number of Weyl points in the Brillouin zone and the topological character of the large Fermi arc in \mote{} is exceptionally sensitive to small changes in the crystal structure as they are observed for samples grown under slightly different conditions. In \fives{} we use an alternative approach to illustrate the persistence of the large Fermi arc for different Weyl point arrangements in \mote. By artificially tuning the strength of the spin-orbit coupling (SOC) (the labels SOC$N$ correspond to SOC increased $N$ times in the calculation), we simulate electronic structures with 8, 4 or 0 Weyl points for a single crystal structure. Clearly, the large Fermi arc is observed in all these calculations. Its presence in experiment~\cite{Belopolski2015,Huang2016arXiv,Deng2016arXiv,Zhou2016arXiv,Shi2016arXiv,Chen2016arXiv} thus cannot serve as a fingerprint to identify \mote{} as Weyl semimetal. 

\begin{figure}[]
\includegraphics[width=6.5cm]{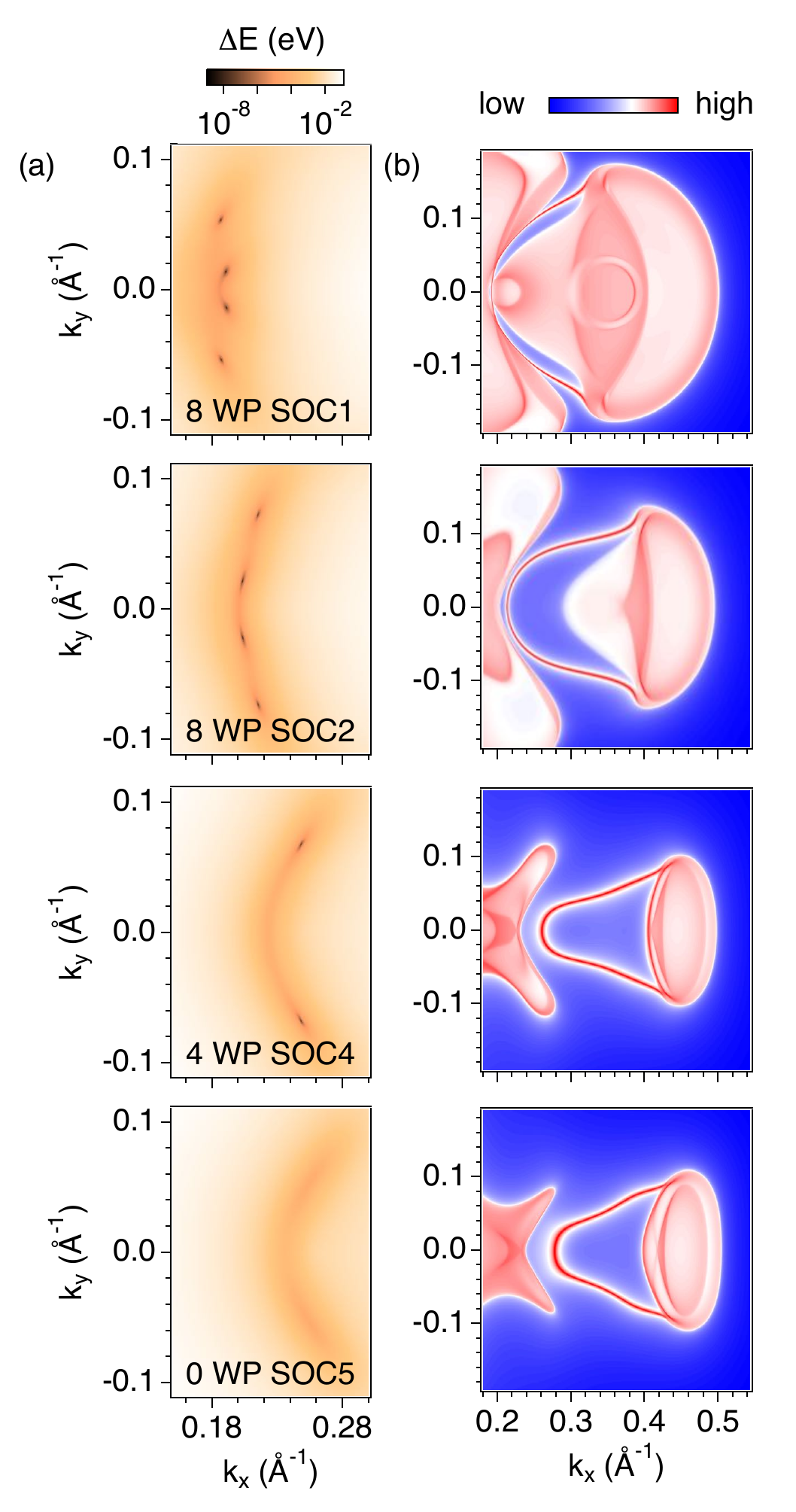} %
\caption{Evolution of the large Fermi arcs for different values of the spin orbit coupling constant, which artificially  modify the band structure  of \mote{} resulting in 8, 4 or 0 weyl points in the $k_{z}=0$ plane. (a) False color plot of the energy gap between valence and conduction bands in the $k_{z}$ = 0 plane illustrating the number and position of Weyl points. (b) Momentum-resolved surface density of states at the Fermi level. }
\label{fs5}
\end{figure}

\begin{figure}[b]
\includegraphics[width=8.6cm]{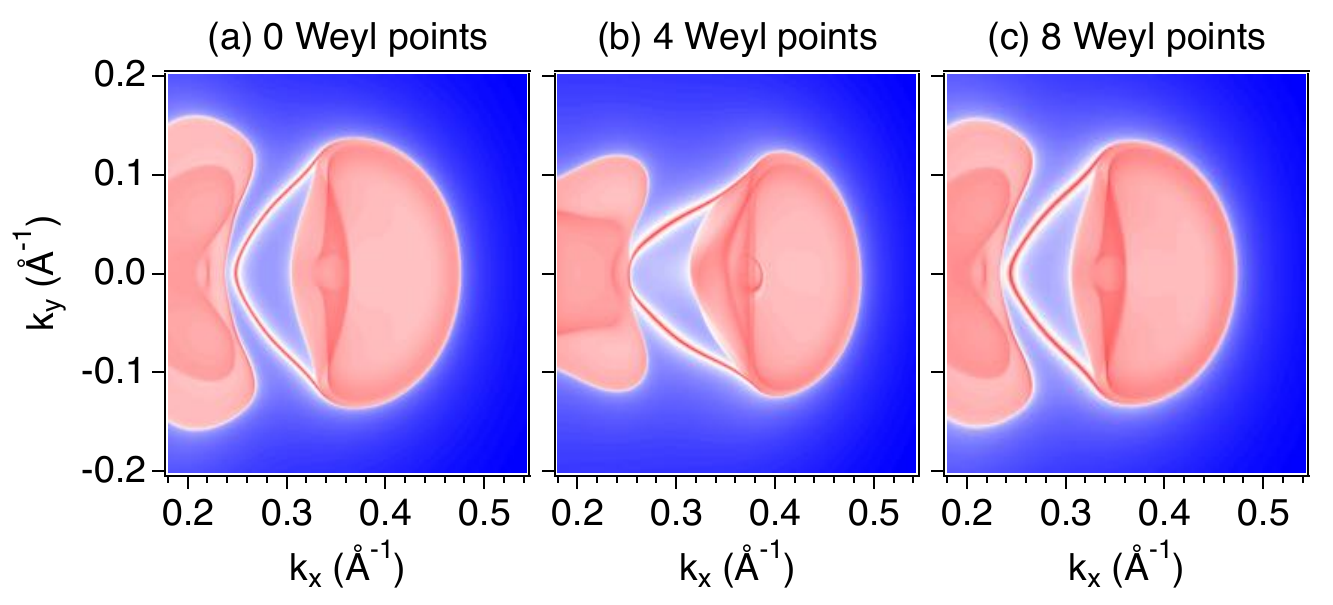} %
\caption{Large Fermi arc of \wte{} for different numbers of Weyl points. (a) Momentum-resolved surface density of states at the Fermi level calculated using the room temperature structure~\cite{Brown1966}, which results in no Weyl points. (b) Momentum-resolved surface density of states at the Fermi level calculated using the low temperature structure (T = 113 K, Ref.~\cite{Mar1992}) and 200 \% spin orbit coupling, which produces only 4 Weyl points. (c)  Momentum-resolved surface density of states at the Fermi level calculated using the low temperature structure (T = 113 K Ref.~\cite{Mar1992}) and 100 \% spin orbit coupling, which produces 8 Weyl points.}
\label{fs6}
\end{figure}

The independence of the large Fermi arc from the Weyl point arrangement is even more striking in \wte{} as shown in \sixs. In panels (a) and (c) we show the momentum resolved surface density of states for experimental crystal structures measured at different temperatures~\cite{Brown1966,Mar1992}. Warming up from 113~K to room temperature, one can readily annihilate the 8 Weyl points found for the low-temperature structure~\cite{Soluyanov2015a}. However, the large Fermi arc remains present and does not even change its dispersion noticeably. Artificially increasing the spin-orbit interaction, we can even simulate a state with a large Fermi arc of similar dispersion and 4 Weyl points in the full 3D Brillouin zone. In this case the large Fermi arc becomes topological, while in the other two calculations (panels (a,c)), it is topologically trivial. Again, we conclude that its observation in experiments~\cite{Bruno2016,Zhou2016arXivWTe2,Wu2016arXiv} does not identify \wte{} as topological Weyl semimetal, as discussed already in Ref.~\cite{Bruno2016}.

The large Fermi arcs in \mote{} and \wte{} remain fully spin-polarized in all our calculations although they are in general topologically trivial except for the calculations with 4 Weyl points. The lifting of the spin degeneracy is a natural consequence of spin-orbit interaction in structures with a broken inversion symmetry and is thus found for all surface and bulk states in the 1T' phase of \mote{} and \wte.  
Hence, the topological character of this and other surface states in \mote{} or \wte{} cannot be deduced from their spin-structure, as it was proposed in Ref.~\cite{Chen2016arXiv} and successfully done for topological insulators~\cite{Hsieh2009}.


%

%
\end{document}